\documentclass[aps,prd,preprintnumbers,superscriptaddress,nofootinbib,notitlepage,floatfix,10pt,twocolumn]{revtex4-2}
\usepackage[pdftex]{graphicx}
\usepackage{bm,latexsym,amsmath,amssymb,amsfonts,mathrsfs,physics,tensor}
\usepackage{color}
\allowdisplaybreaks[1]
\usepackage[pdftex,colorlinks=true,linkcolor=blue,citecolor=cyan,backref=page]{hyperref}
\newcommand*{\D}{\mathrm{d}}

\newcommand*{\tF}{\widetilde{F}}
\newcommand*{\gc}{\mathsf{g}}
\newcommand*{\expara}{\lambdabar}
\newcommand*{\ups}{\Xi}

\begin{document}
\title{Spinoptics in the presence of axion-like particles in curved spacetime}
%
\author{Tomoki~Takeuchi}
\email[Email: ]{t.takeuchi@rikkyo.ac.jp}
\affiliation{Department of Physics, Rikkyo University, Toshima, Tokyo 171-8501, Japan}
\author{Tsutomu~Kobayashi}
\email[Email: ]{tsutomu@rikkyo.ac.jp}
\affiliation{Department of Physics, Rikkyo University, Toshima, Tokyo 171-8501, Japan}
%
\begin{abstract}
We study the propagation of high-frequency electromagnetic waves coupled to an axionic scalar field in curved spacetime.
By applying the effective action approach, we go beyond the standard geometric optics limit and derive spinoptics equations for the axion--Maxwell theory that are valid for arbitrary curved spacetimes and arbitrary axion profiles.
These equations allow us to analyze helicity-dependent corrections to photon trajectories arising from the photon's interactions with both the axion field and spacetime curvature.
We present representative examples to demonstrate how light trajectories deviate from null geodesics due to spinoptics effects in the presence of an axion field.
\end{abstract}
\preprint{RUP-26-16}
\maketitle

\section{Introduction}\label{sec:intro}

Electromagnetic (and gravitational) waves deliver valuable information to us in astronomical observations.
If their wavelengths are sufficiently short compared to other characteristic length scales in the system considered, one can use the standard geometric optics approximation~\cite{Misner:1973prb, Maggiore:2007ulw}.
The notion of rays then emerges, and the trajectories of a photon (and a graviton) are described by null geodesics in a curved spacetime.
A ray is deflected when passing by a massive object such as a black hole.
Gravitational time dilation also occurs when a ray passes through a gravitational field.
These effects help us to deduce the mass distribution in the region through which a ray travels.
Ray trajectories thus play a crucial role in astrophysics.

While these are the effects of spacetime curvature on ray trajectories, the polarization vector is also affected by the curvature: the polarization plane of linearly polarized light rotates when passing near a rotating body such as a Kerr black hole.
This is the gravitational Faraday effect, which has a long history~\cite{Balazs:1958nl, Plebanski:1959ff, Dehnen:1973jw, Ishihara:1987dv, Nouri-Zonoz:1999jls, Kopeikin:2001dz, Shoom:2020zhr, Shoom:2024zep}.

There is another interesting effect called the gravitational spin Hall effect that arises from the interaction between the spin of a photon and spacetime curvature~\cite{Gosselin:2006wp, Frolov:2011mh, Dolan:2018ydp, Shoom:2020zhr, Shoom:2024zep} (see Refs.~\cite{Oancea:2019pgm, Andersson:2023bvw} for reviews and Refs.~\cite{Yoo:2012vv, Yamamoto:2017gla, Andersson:2020gsj, Dahal:2021qel, Kubota:2023dlz, Kubota:2024zkv, Frolov:2024qow} for the spin Hall effect of gravitational waves).
The gravitational spin Hall effect describes how the polarization vector affects ray trajectories, yielding helicity-dependent deviations from null geodesics.
To take into account this effect, we need to go beyond the standard geometric optics limit.
Several different approaches have been developed to extend the standard geometric optics~\cite{Frolov:2011mh, Yoo:2012vv, Dolan:2017zgu, Dolan:2018ydp, Oancea:2020khc, Nishida:2026bzu}.
Such an extension of geometric optics is called spinoptics.
Among different approaches, the effective action method developed recently by Frolov~\cite{Frolov:2024ebe} is particularly simple and transparent, allowing one to derive spinoptics equations systematically in an arbitrary curved spacetime.
The resultant spinoptics equations have been solved in Schwarzschild~\cite {Frolov:2024olb, Murk:2024qgj}, Kerr~\cite{Dahal:2023ncl, Frolov:2025bva}, and hairy black-hole~\cite{Alves:2026jyc} spacetimes, and thereby the corrections to null geodesics have been evaluated qualitatively and quantitatively.

Considerable attention has been paid to axions in recent studies of beyond the Standard Model physics.
It was originally proposed in the context of QCD as a solution to the strong CP problem~\cite{Peccei:1977hh, Peccei:1977ur, Weinberg:1977ma, Wilczek:1977pj, Preskill:1982cy, Dine:1981rt} (see Refs.~\cite{Kim:2008hd, DiLuzio:2020wdo} for reviews).
Light particles with similar properties have also been considered in different contexts, such as low-energy effective theories of string theory.
These light particles are called axion-like particles (ALPs)~\cite{Witten:1984dg, Arvanitaki:2009fg}.
The axion is one of the leading dark matter candidates~\cite{Preskill:1982cy, Abbott:1982af, Dine:1982ah, Hui:2016ltb}, and it could also be dark energy~\cite{Frieman:1995pm, Kim:2003}.

The crucial property of the ALP field is its coupling to photons described by the interaction Lagrangian
\begin{align}
    \mathcal{L}_{\textrm{int}}=-\frac{\gc}{4}\phi F_{\mu\nu}\tF^{\mu\nu},
    \label{int-L-axion-photon}
\end{align}
where $\gc$ is the coupling constant, $\phi$ is the ALP field, $F_{\mu\nu}$ is the Faraday tensor, and $\tF^{\mu\nu}$ is its dual.
It is important to understand in depth the propagation of electromagnetic waves in the presence of the interaction~\eqref{int-L-axion-photon}.
It is known that this coupling induces rotation of the polarization plane~\cite{Harari:1992ea, Fedderke:2019ajk} that is similar to, but fundamentally different from, the standard Faraday rotation induced by a magnetic field~\cite{Landau1984v2}.
This phenomenon has been of considerable interest in particular in the context of cosmology (see, e.g., Refs.~\cite{Carroll:1998zi, Lue:1998mq, Feng:2006dp, Liu:2006uh, Fujita:2020ecn, Komatsu:2022nvu}).
The next question is how the axion--photon coupling~\eqref{int-L-axion-photon} affects ray trajectories associated with high-frequency electromagnetic waves.
Though some papers suggested helicity-dependent bending of light by ALPs~\cite{Plascencia:2017kca, Chway:2019prm, McDonald:2019wou}, the question has been addressed in the geometric optics limit, showing that in arbitrary curved spacetime ray trajectories are still given by null geodesics, and the only effect of the axion--photon interaction~\eqref{int-L-axion-photon} is the aforementioned rotation of the polarization plane (to all orders in $\gc$)~\cite{Blas:2019qqp, Fedderke:2019ajk, Schwarz:2020jjh}.

It is then natural to consider the possibility of the helicity-dependent bending of light beyond the geometric optics limit in the presence of the axion--photon interaction.
In this work, we extend the previous studies~\cite{Blas:2019qqp, Fedderke:2019ajk, Schwarz:2020jjh} and develop a spinoptics approximation in the axion--Maxwell theory in arbitrary curved spacetime and for an arbitrary axion-field configuration.
To do so, we use the effective action approach employed in the case of Maxwell theory in a curved background~\cite{Frolov:2024ebe}.
We also present representative examples of solutions to the derived spinoptics equations to understand the effects of ALPs on ray trajectories beyond the geometric optics limit.

This paper is organized as follows.
In the next section, we derive the spinoptics equations in the presence of ALPs in curved spacetime.
We then solve the spinoptics equations for two representative examples of spacetimes and axion-field configurations and demonstrate the behavior of scattering of polarized light in Sec.~\ref{sec:examples}.
In Sec.~\ref{sec:photon_sphere}, we discuss the consequences of the spinoptics approximation on the photon sphere of a black hole dressed with an ALP field.
Finally, we draw our conclusions in Sec.~\ref{sec:conclusions}.

Our convention for the Riemann tensor follows that of Wald~\cite{Wald:1984rg}:
$R_{\mu\nu\rho}^{~~~~\sigma}=\partial_\nu\Gamma^\sigma_{\mu\rho}-\partial_\mu\Gamma^\sigma_{\nu\rho}+\Gamma^\alpha_{\mu\rho}\Gamma_{\alpha\nu}^\sigma-\Gamma^\alpha_{\nu\rho}\Gamma_{\alpha\mu}^\sigma$.

\section{Spinoptics equations in axion--Maxwell theory}\label{sec:basic-equations}

Let $\lambdabar$ be the characteristic (reduced) wavelength of the electromagnetic waves.
We assume that $\lambdabar$ is much smaller than the characteristic length scales over which the gravitational field and the axion field vary:
$\lambdabar/L_g\ll 1$, $\lambdabar/L_\phi\ll 1$,
where $\partial_\mu g_{\alpha\beta}\sim 1/L_g$ and $\partial_\mu\phi\sim \phi/L_\phi$.

\subsection{Effective action}\label{subsec:EA}

The Lagrangian for the axion--Maxwell theory in an arbitrary curved spacetime is given by 
\begin{align}
        \mathcal{L}=-\frac{1}{4}
        \left(
            F_{\mu\nu}F^{\mu\nu}
            +\gc \phi F_{\mu\nu}\tF^{\mu\nu}
        \right),
\end{align}
where $F_{\mu\nu}=\partial_\mu A_\nu -\partial_\nu A_\mu$ and
\begin{align}
        \tF^{\mu\nu}=\frac{1}{2}\varepsilon^{\mu\nu\rho\sigma}F_{\rho\sigma}.
\end{align}
Here, $\varepsilon^{\mu\nu\rho\sigma}:=\epsilon^{\mu\nu\rho\sigma}/\sqrt{-g}$
is the totally antisymmetric tensor with $\epsilon^{0123}=1$,
and $\gc$ is a constant.
In this paper, the axion field $\phi$ is also treated as a fixed background.

To derive the spinoptics equations in the presence of ALPs, we closely follow the elegant formulation by Frolov~\cite{Frolov:2024ebe}.
Let us consider a complex field action 
\begin{align}
        I=-\frac{1}{4}\int\D^4x\sqrt{-g}
        \left[\bar F_{\mu\nu}F^{\mu\nu}+\gc \phi \bar F_{\mu\nu}\tF^{\mu\nu}\right],
        \label{eq:complex-action}
\end{align}
where an overbar denotes complex conjugation.
We look for a solution of the form
\begin{align}
        A_\mu=AM_\mu e^{iS/\expara},\label{eq:ansatz-for-A}
\end{align}
with $A$ and $S$ being real scalar functions.
The complex vector $M_\mu$ satisfies
\begin{align}
        &M_\mu\bar M^\mu=1,
        \qquad 
        M_\mu M^\mu = \bar M_\mu \bar M^\mu =0,
        \notag \\ &
        M^\mu S_{,\mu}=0.
        \label{eq:constraints-on-M}
\end{align}
The first constraint is just a normalization.
The second one implies that the solution we look for corresponds to circular polarized light.
The constraint $M^\mu S_{,\mu}=0$ amounts to a $U(1)$ gauge choice.
There still remains a residual gauge degree of freedom, which will be fixed later.

We substitute the ansatz~\eqref{eq:ansatz-for-A} to the action~\eqref{eq:complex-action},
and retain next-to-leading order terms in the $\expara$ expansion.
We have 
\begin{align}
        F_{\mu\nu}&=\frac{i}{\expara}A\left(
                S_{,\mu}M_\nu -S_{,\nu}M_\mu
        \right)e^{iS/\expara}
        \notag \\ &\quad 
        +
        \left[
                \nabla_{\mu}\left(AM_\nu\right)-\nabla_{\nu}\left(AM_\mu\right)
        \right]e^{iS/\expara}.
\end{align}
Up to a total derivative, we obtain
\begin{align}
        \phi \bar F_{\mu\nu}\tF^{\mu\nu}=-\frac{2i}{\expara}\left[
        \varepsilon^{\mu\nu\rho\sigma}\nabla_\nu\phi\cdot A^2\bar M_\rho M_\sigma S_{,\mu}
        +\mathcal{O}(\expara)
        \right].
\end{align}
Thus, to first subleading order in $\expara$, the action reduces to
\begin{align}
        I[A,M^\mu,S]=\frac{1}{\expara^2}\int \D^4x\sqrt{-g}A^2
        \left[
                -\frac{1}{2}(\nabla S)^2+\expara B^\mu S_{,\mu}
        \right],
        \label{eq:eff-act-00}
\end{align}
where 
\begin{align}
        B^\mu:=i\bar M^\nu \nabla^\mu M_\nu +i\frac{\gc}{2} \varepsilon^{\mu\nu\alpha\beta}
        \phi_{,\nu}\bar M_\alpha M_\beta.
        \label{eq:def-B}
\end{align}
The apparent form of the action~\eqref{eq:eff-act-00} is the same as that given in Ref.~\cite{Frolov:2024ebe}.
At this point, the difference is the additional contribution in $B^\mu$ (the second term in Eq.~\eqref{eq:def-B}) from the axion field.
Note that $B^\mu$ is real.

We drop the overall factor $1/\expara^2$ and include the constraints to enforce Eq.~\eqref{eq:constraints-on-M}, arriving at the effective action
\begin{align}
    I'=\int \D^4x\sqrt{-g}
        \left\{
                A^2
                \left[
                -\frac{1}{2}(\nabla S)^2+\expara B^\mu S_{,\mu}
                \right]
                +C
        \right\},
        \label{effective-action}
\end{align}
with
\begin{align}
    C&=\frac{\bar u_1}{2} M_\mu M^\mu 
    +\frac{u_1}{2}\bar M_\mu \bar M^\mu 
    +u_2\left(M_\mu \bar M^\mu -1\right)
    \notag \\ & \quad 
    +\bar u_3M^\mu S_{,\mu}+u_3\bar M^\mu S_{,\mu}.
\end{align}
Varying $I$ with respect to $A$, $\bar M^\mu$, and $S$, we obtain the corresponding equations of motion as
\begin{align}
        &H:=
        \frac{1}{2}(\nabla S)^2-\expara B^\mu S_{,\mu}
        =0,
        \label{eom:Hamiltonian}
        \\
        &\expara iA^2\left(
                \nabla^\mu M^\nu+\frac{1}{2}
                \gc \varepsilon^{\mu\alpha\nu\beta}\phi_{,\alpha}
                M_\beta
        \right)S_{,\mu}
        \notag \\ &
        +u_1\bar M^\nu+u_2 M^\nu+u_3 S^{;\nu}
        =0,\label{eom:barm}
        \\
        &
        \nabla_\mu J^\mu=0, \label{eom:photonnum}
\end{align}
where
\begin{align}
        J^\mu:=
        A^2\left(\nabla^\mu S-\expara B^\mu\right)
        -\bar u_3M^\mu -u_3\bar M^\mu.
        \label{def:J-mu}
\end{align}

Equation~\eqref{eom:barm} is the transport equation for the polarization vector.
By multiplying it by $M_\nu$ and using Eq.~\eqref{eq:constraints-on-M} and $\varepsilon^{\mu\alpha\nu\beta}M_\nu M_\beta=0$, one immediately obtains $u_1=0$.
Equation~\eqref{eq:constraints-on-M} is invariant under the following transformation,
\begin{align}
    M^\mu \to e^{i \psi }M^\mu, \qquad \bar{M}^\mu \to e^{-i\psi}\bar{M}^\mu. 
\end{align}
Under this transformation, $B_\mu$ transforms as $B_\mu \to B_\mu - \partial_\mu \psi$.
Using this freedom, one can set 
\begin{align}
    B^\mu S_{,\mu} = 0.\label{eq:gauge-fix-B}
\end{align}
Multiplying Eq.~\eqref{eom:barm} by $\bar{M}^\nu$ and using~\eqref{eq:constraints-on-M} and~\eqref{eq:gauge-fix-B}, we obtain $u_2=0$.
Thus, Eq.~\eqref{eom:barm} can be written as
\begin{align}
    &\expara iA^2\left(
                \nabla^\mu M^\nu+\frac{1}{2}
                \gc \varepsilon^{\mu\alpha\nu\beta}\phi_{,\alpha}
                M_\beta
        \right)S_{,\mu}
        \notag \\ &
        +u_3 S^{;\nu}
        =0.
        \label{eq:transport-polarization-vec}
\end{align}
The Lagrange multiplier $u_3$ is left undetermined, as there is the aforementioned residual gauge degree of freedom.
One can see, however, that $u_3=\mathcal{O}(\expara)$ or higher.

Equation~\eqref{eom:Hamiltonian} takes the form of the Hamilton--Jacobi equation, defining a local dispersion relation.
The characteristic curves of the Hamilton--Jacobi equation~\eqref{eom:Hamiltonian}, which define ray trajectories (even beyond the geometric optics limit),\footnote{For a detailed discussion, see Ref.~\cite{Perlick:2000jx}.} coincide with the solutions to the Hamiltonian equations for $H$ obtained by setting $\nabla_\mu S=p_\mu$:
\begin{align}
    H(x,p)=\frac{1}{2}g^{\mu\nu}p_\mu p_\nu-\expara B^\mu p_\mu.
    \label{Ham-eq-x-p}
\end{align}
The Hamiltonian equations are given by
\begin{align}
    \frac{\D x^\mu}{\D\lambda}&=p^\mu - \expara B^\mu 
    \label{Ham-eq-1}
    \\ 
    \frac{\D p_\mu}{\D\lambda}&=\frac{1}{2}\partial _\mu g^{\alpha \beta} p_\alpha p_\beta - \expara \partial_\mu B^\alpha  p_\alpha,
    \label{Ham-eq-2}
\end{align}
where $\lambda$ is the parameter of a curve: $x^\mu=x^\mu(\lambda)$.
The ray trajectories are derived by solving these equations.

\subsection{Null tetrad}

Let $l^\mu$ be a tangent vector to the ray:
\begin{align}
    l^\mu=\frac{\D x^\mu}{\D \lambda}=p^\mu-\expara B^\mu.\label{eq:tangentvector}
\end{align}
From Eqs.~\eqref{eom:Hamiltonian} and~\eqref{Ham-eq-x-p}, we see that $l^\mu$ is a null vector to first order in $\expara$:
\begin{align}
    l^\mu l_\mu = {\mathcal O}(\expara ^2).
\end{align}
This conclusion holds even in the presence of the axion-dependent term in $B^\mu$.
Using the Hamiltonian equations~\eqref{Ham-eq-1} and~\eqref{Ham-eq-2}, we obtain
\begin{align}
    D l^\mu = \expara \kappa^\mu,
    \qquad
    \kappa_\mu := \left(\nabla_\mu B_\nu - \nabla_\nu B_\mu\right)l^\nu, \label{eom:light rays}
\end{align}
with $D:=l^\nu \nabla_\nu$.
Note that $l^\mu \kappa_\mu = 0$.
In the geometric optics limit, $\expara\to 0$, the equation reduces to a geodesic equation.
The right-hand side describes how a trajectory $x^\mu(\lambda)$ deviates from a geodesic at first order in $\expara$.

Now we consider a congruence of null rays and a complex null tetrad $(l^\mu,n^\mu,m^\mu,\bar m^\mu)$ associated with it.
These vectors are normalized as follows,
\begin{align}
    l^\mu n_\mu = - m^\mu \bar m_\mu = -1, \label{eq:normalization}
\end{align}
and all other inner products are zero.
The complex null tetrad admits the following freedom that preserves the normalization conditions:
\begin{align}
    \textbf{1.}& ~~l^\mu \to \gamma l^\mu, \qquad n^\mu \to \gamma^{-1}n^\mu. 
    \label{tetrad-transformation-1}\\
    \textbf{2.}&~~m^\mu \to e^{i\psi}m^\mu , \qquad  \bar m^\mu \to e^{-i \psi}\bar m^\mu . 
    \label{tetrad-transformation-2}\\
    \textbf{3.}&~~l^\mu \to l^\mu, \qquad 
    m^\mu \to m^\mu + \alpha l^\mu , \bar m^\mu \to \bar m^\mu +\bar \alpha l^\mu, 
    \notag \\
    &~~n^\mu \to n^\mu  + \bar \alpha  m^\mu + \alpha \bar m^\mu + \alpha \bar \alpha l^\mu.
    \label{tetrad-transformation-3}
    \\
    \textbf{4.}&~~m^\mu \to \bar m^\mu \qquad \bar m^\mu \to m^\mu,
    \label{tetrad-transformation-4}
\end{align}
where $\gamma$ and $\psi$ are real functions and $\alpha $ is a complex function. 
Using the normalization conditions, it is easy to show that
$\varepsilon_{\mu\nu\alpha\beta}l^\mu n^\nu m^\alpha \bar{m}^\beta=i\sigma$,
where $\sigma=\pm1$ is the helicity parameter.
A complex null tetrad with $\sigma=+1$ ($\sigma=-1$) is said to be right-handed (left-handed).
The transformation~\eqref{tetrad-transformation-4} flips the helicity.
In the derivation of the spinoptics equations below, we will proceed by taking $\sigma=+1$, and restore $\sigma$ in the final step.

Since $l^\mu\kappa_\mu = 0$, one can write
\begin{align}
    \kappa^\mu=\kappa_l l^\mu+\bar \kappa m^\mu +\kappa \bar m^\mu,
\end{align}
where $\kappa_l$ is a real function and $\kappa$ is a complex one.

Let us specify how $m^\mu$ and $n^\mu$ are transported along the null ray.
In the case of Maxwell theory, it is appropriate to require that they are parallel-transported in the $\expara\to 0$ limit and write the transport equations in the form of
$Dm^\mu=\mathcal{O}(\expara)$ and $Dn^\mu=\mathcal{O}(\expara)$~\cite{Frolov:2024ebe}.
In the presence of the axion field, it is more convenient to modify the leading order part as
\begin{align}
    (\mathcal{D} m)^\mu = \expara Z^\mu,\qquad 
    (\mathcal{D} n)^\mu = \expara N^\mu,
\end{align}
where
\begin{align}
        (\mathcal{D}v)^\alpha:=D v^\alpha
        -\frac{\gc}{2}\varepsilon^{\mu\nu\lambda\alpha}l_\mu \phi_{,\nu}v_\lambda,
        \label{def:cal-D}
\end{align}
so that $m^\mu$ and $n^\mu$ are transported under the influence of the axion field.
The reason for this modification is to be clarified later.
Note that
\begin{align}
    D(l^\mu n_\mu) &= l^\mu \left[
    (\mathcal{D}n)_\mu +\frac{\gc}{2}\varepsilon_{\nu\lambda\alpha\mu}l^\nu \phi^{;\lambda}n^\alpha
    \right]+n_\mu Dl^\mu
    \notag \\ 
    &=\expara\left(l^\mu N_\mu-\kappa_l\right),
    \\ 
    D(l^\mu m_\mu)&=\expara\left(l^\mu Z_\mu+\kappa\right),
    \\ 
    D(n^\mu m_\mu)&=
    n^\mu \left[
    (\mathcal{D}m)_\mu +\frac{\gc}{2}\varepsilon_{\nu\lambda\alpha\mu}l^\nu \phi^{;\lambda}m^\alpha
    \right]
    \notag \\ & \quad  
    +m^\mu \left[
    (\mathcal{D}n)_\mu +\frac{\gc}{2}\varepsilon_{\nu\lambda\alpha\mu}l^\nu \phi^{;\lambda}n^\alpha
    \right]
    \notag \\ &
    =
    \expara\left(n^\mu Z_\mu + m_\mu N^\mu\right),
    \\ 
    D(n^\mu n_\mu)&=2\expara N^\mu n_\mu,
    \\ 
    D(m^\mu m_\mu)&=2\expara Z^\mu m_\mu, 
    \\ 
    D(m^\mu \bar m_\mu)&=\expara\left(m^\mu \bar Z_\mu+\bar m^\mu Z_\mu\right).
\end{align}
In order for the normalization conditions to be maintained along the null ray, it is required that
\begin{align}
    &n_\mu N^\mu = m_\mu Z^\mu = \bar m_\mu \bar Z^\mu =0,
    \\ 
    &l^\mu N_\mu =\kappa_l,
    \\ 
    &l^\mu Z_\mu =-\kappa,
    \\ 
    &n^\mu Z_\mu +m_\mu N^\mu =0,
    \\ 
    &m^\mu \bar Z_\mu+\bar m^\mu Z_\mu =0.
\end{align}
Therefore, $Z^\mu$ and $N^\mu$ must be of the form
\begin{align}
    Z^\mu&=Z_l l^\mu+\kappa n^\mu +iZ m^\mu,
    \\ 
    N^\mu&=-\kappa_l n^\mu+\bar Z_l m^\mu + Z_l \bar m^\mu,
\end{align}
where $Z$ is real and $Z_l$ is complex.
However, by the use of the transformations~\eqref{tetrad-transformation-2} and~\eqref{tetrad-transformation-3} one can set $Z=Z_l=0$~\cite{Frolov:2024ebe}.
Thus, one can construct a complex null tetrad satisfying
\begin{align}
    (\mathcal{D}m)^\mu &= \expara \kappa n^\mu,
    \label{eq:transport-m}
    \\ 
    (\mathcal{D}n)^\mu &=-\expara \kappa_l n^\mu.
    \label{eq:transport-n}
\end{align}
A complex null tetrad $(l^\mu, n^\mu, m^\mu ,\bar m^\mu )$ chosen in this way plays a crucial role in formulating the spinoptics equations.

\subsection{Spinoptics equations}

We now write the equations of motion obtained in Sec.~\ref{subsec:EA} in terms of a null tetrad $(l^\mu, n^\mu, m^\mu, \bar m^\mu)$.
The polarization vector $M^\alpha$ can be expressed consistently as
\begin{align}
    M^\alpha  = m^\alpha  + \expara \mu^\alpha,
    \label{eq:ansatz-for-M}
\end{align}
where, although the constraints~\eqref{eq:constraints-on-M} allow for a component along $l^\mu$ at leading order, $M^\mu=m^\alpha + a l^\alpha+\expara \mu^\alpha$, the residual $U(1)$ gauge degree of freedom can be fixed so that there is no such component.
It follows from the definition of $B^\mu $ that, to first order in $\expara$,
\begin{align}
    B^\mu &= b^\mu + \expara \beta^\mu,
    \\
    b^\mu &= i m^\nu \nabla ^\mu m_\nu + i\frac{\gc}{2}\varepsilon^{\mu\nu\alpha\beta}\phi_{,\nu}\bar m_\alpha m_\beta,
    \\ 
    \beta^\mu &=i ( \bar \mu ^\nu \nabla^\mu m_\nu+ \bar m^\nu \nabla^\mu \mu _\nu )
    \notag \\ & \quad 
    + i\frac{\gc}{2}\varepsilon^{\mu\nu\alpha\beta}\phi_{,\nu}( \bar\mu_\alpha m_\beta+ \bar m_\alpha \mu_\beta ).\label{B1+2}
\end{align}
Using $S_{,\mu}=p_{\mu}=l_\mu + \expara b_\mu+\mathcal{O}(\expara^2)$, we see from Eq.~\eqref{eq:transport-polarization-vec} that
\begin{align}
    \expara i A^2(\mathcal{D}m)^\mu
    +u_3l^\mu=\mathcal{O}(\expara^2).
\end{align}
Since $(\mathcal{D}m)^\mu=\mathcal{O}(\expara)$, this equation implies that $u_3=\mathcal{O}(\expara^2)$.
The axion-corrected transport equation~\eqref{eq:transport-m} and the gauge choice to write $M^\alpha$ in the form~\eqref{eq:ansatz-for-M} are crucial here.

Since the right-hand side of Eq.~\eqref{eom:light rays} already carries a factor of $\expara$, it is sufficient to evaluate $B_\mu$ at leading order, replacing it with $b_\mu$: $\kappa_\mu=(\nabla_\mu b_\nu-\nabla_\nu b_\mu)l^\nu$.
We can write $\kappa_\mu$ explicitly as
\begin{align}
    \kappa_\mu &= 
        i\left[
                R_{\mu\nu\alpha\beta}+\frac{\gc^2}{4}
                \left(
                        g_{\mu\beta}\phi_{,\nu}\phi_{,\alpha}
        -g_{\mu\alpha}\phi_{,\nu}\phi_{,\beta}
                \right)
        \right]\bar m^\alpha m^\beta l^\nu
        \notag \\ & \quad 
        -\frac{\gc}{2}\left(
                m_\mu l^\alpha \bar m^\beta
                +\bar m_\mu l^\alpha m^\beta
                -2l_\mu l^\alpha n^\beta
        \right)\nabla_\alpha\nabla_\beta\phi.
        \label{kappa-explicit}
\end{align}
To derive this expression, we used
\begin{align}
    g_{\mu\nu}=-l_\mu n_\nu-l_\nu n_\mu+m_\mu\bar m_\nu+m_\nu\bar m_\mu,
\end{align}
and 
\begin{align}
    0&=l^\alpha \nabla_\mu\left(-l_\alpha n^\nu-n_\alpha l^\nu
    +m_\alpha \bar m^\nu+\bar m_\alpha m^\nu
    \right)
    \notag \\ 
    &=\nabla_\mu l^\nu -l^\nu l^\alpha \nabla_\mu n_\alpha
    +\bar m^\nu l^\alpha\nabla_\mu m_\alpha+m^\nu l^\alpha\nabla_\mu \bar m_\alpha.
\end{align}
From Eq.~\eqref{kappa-explicit}, we obtain
\begin{align}
        \kappa_l&=iR_{\mu\nu\alpha\beta} l^\mu n^\nu\bar m^\alpha m^\beta 
        +\gc l^\mu n^\nu\nabla_\mu\nabla_\nu\phi,
        \\
        \kappa&=
        iR_{\mu\nu\alpha\beta}m^\mu l^\nu \bar m^\alpha m^\beta
        +ie^{-i\gc \phi/2} m^\mu l^\nu \nabla_\mu\nabla_\nu 
        e^{i\gc \phi/2}.
\end{align}

Having shown that $u_3=\mathcal{O}(\expara^2)$, Eqs.~\eqref{eom:photonnum} and~\eqref{def:J-mu} now reduce to
\begin{align}
    \nabla_\mu\left(A^2 l^\mu\right)=\mathcal{O}(\expara^2).
\end{align}
Thus, the number of photons is conserved to first order in $\expara$ even in the presence of the axion field.

To derive the transport equation for $\mu^\alpha$, we need $\mathcal{O}(\expara^2)$ terms in Eq.~\eqref{eq:transport-polarization-vec}.
However, our effective action~\eqref{effective-action} is valid up to $\mathcal{O}(\expara)$, and hence we cannot determine $\mu^\alpha$ within the validity of the adopted approximation.
This is not a fatal drawback for our purpose, because we only need the leading order part in $M^\alpha$ to determine $\mathcal{O}(\expara)$ corrections to a ray trajectory.

Restoring the helicity dependence, the spinoptics equations in the axion--Maxwell theory are summarized as
\begin{align}
    Dl^\mu&=\expara\sigma \kappa^\mu,\label{SOeqs-l}
    \\ 
    (\mathcal{D} m)^\mu&=\expara\sigma \kappa n^\mu,\label{SOeqs-m}
    \\ 
    (\mathcal{D} n)^\mu &=-\expara\sigma\kappa_ln^\mu,\label{SOeqs-n}
    \\ 
    \nabla_\mu\left(A^2l^\mu\right)&=0.
\end{align}
We treat the right-hand side of the spinoptics equations~\eqref{SOeqs-l}--\eqref{SOeqs-n} as perturbations of order $\expara$.
The next step is therefore to construct a complex null tetrad $(\ell_0^\mu,n_0^\mu,m_0^\mu,\bar m_0^\mu)$ satisfying the transport equations in the $\expara\to 0$ limit:
\begin{align}
    D_0l^\mu_0&=0,
    \\ 
    (\mathcal{D}_0m_0)^\mu=(\mathcal{D}_0n_0)^\mu&=0,
    \label{eq:calDnm0}
\end{align}
where $D_0:=l_0^\mu \nabla_\mu$ and
$(\mathcal{D}_0v)^\alpha:=D_0v^\alpha-(\gc/2)\varepsilon^{\mu\nu\lambda\alpha}l_{0\mu}\phi_{,\nu}v_\lambda$.
Evaluating the right-hand side of Eq.~\eqref{SOeqs-l} using $(\ell_0^\mu,n_0^\mu,m_0^\mu,\bar m_0^\mu)$, one can obtain $\mathcal{O}(\expara)$ deviations from a null geodesic.

\subsection{Parallel-transported basis}

The ``axion-corrected'' tetrad $(\ell_0^\mu,n_0^\mu,m_0^\mu,\bar m_0^\mu)$ can be obtained from another complex null tetrad $(l^\mu_0,\mathsf{n}_0^\mu,\mathsf{m}_0^\mu,\bar{\mathsf{m}}_0^\mu)$ parallel-propagated along the null geodesic,
\begin{align}
    D_0\mathsf{m}_0^\mu=D_0\mathsf{n}_0^\mu=0.
    \label{eq:para-n-m}
\end{align}
Once a parallel-propagated null tetrad is obtained, we define $m^\mu_0$ and $n^\mu_0$ by
\begin{align}
        m_0^\mu &=e^{-i \gc \phi/2}
        \left(\mathsf{m}_0^\mu +\ups l^\mu_0\right),
        \label{eq:def-m0}
        \\
        n_0^\mu&=\mathsf{n}_0^\mu
        +\bar{\ups}\mathsf{m}_0^\mu
        +\ups\bar{\mathsf{m}}_0^\mu
        +\ups\bar{\ups} l_0^\mu,
        \label{eq:def-n0}
\end{align}
where
$\ups$ is a complex function that solves
\begin{align}
        D_0\left(e^{-i \gc \phi/2}\ups\right)=
        -\mathsf{m}_0^\nu\left(e^{-i \gc \phi/2}\right)_{,\nu}.
        \label{eq:Upsilon}
\end{align}
The general solution to this equation is given by
\begin{align}
    \ups(\lambda)=-e^{i\gc\phi(\lambda)/2}\left[\int^\lambda_{-\infty}
    \mathsf{m}_0^\nu\left(e^{-i \gc \phi/2}\right)_{,\nu}\D \lambda'
    -\ups_{0}\right],
    \label{soln:Upsilon}
\end{align}
where $\ups_{0}$ is an integration constant.
The choice of $\ups_0$ is discussed later.
It is straightforward to show that $m_0^\mu$ and $n_0^\mu$ obtained from Eqs.~\eqref{eq:def-m0} and~\eqref{eq:def-n0} satisfy Eq.~\eqref{eq:calDnm0}.
Thus, instead of solving Eq.~\eqref{eq:calDnm0}, one may solve Eqs.~\eqref{eq:para-n-m} and~\eqref{eq:Upsilon}.
Note that the transport equations~\eqref{eq:para-n-m} depend only on the metric and is independent of the configuration of the axion field.
The influence of the axion field can be separated in this way.
This is useful in the case where one is able to construct a complex null tetrad $(l^\mu_0,\mathsf{n}_0^\mu,\mathsf{m}_0^\mu,\bar{\mathsf{m}}_0^\mu)$ relatively easily.
Furthermore, although the spinoptics equations in the previous subsection look simpler when expressed in terms of the original axion-corrected basis, it turns out that the parallel-transported basis is more convenient for the purpose of formulating the deviation equations for a ray trajectory.

Let us comment on the results in the geometric optics limit obtained in Ref.~\cite{Schwarz:2020jjh}.
Our transport equation in the $\expara\to 0$ limit, $(\mathcal{D}_0m_0)^\mu=0$, agrees with the corresponding equation in Ref.~\cite{Schwarz:2020jjh}.
However, in the expression for $m^\mu_0$ in terms of a parallel-transported basis, we have the contribution along $l^\mu_0$, which seems to be missing in Ref.~\cite{Schwarz:2020jjh}.
Though at the level of the solution to $(\mathcal{D}_0m_0)^\mu=0$ this contribution is necessary, it does not change the physical conclusions of Ref.~\cite{Schwarz:2020jjh}.

The following expression in terms of the parallel-transported basis is useful for later calculations:
\begin{align}
        e^{i\gc\phi/2}\kappa &= iR_{\mu\nu\alpha\beta}\mathsf{m}_0^\mu l_0^\nu
        \left(
                \bar{\mathsf{m}}_0^\alpha \mathsf{m}_0^\beta
                +\ups\bar{\mathsf{m}}_0^\alpha l_0^\beta
                -\bar\ups \mathsf{m}_0^\alpha l_0^\beta
        \right)
        \notag \\ & \quad 
        +ie^{-i\gc\phi/2}\left(\mathsf{m}_0^\mu+\ups l_0^\mu\right)l_0^\nu 
        \nabla_\mu\nabla_\nu e^{i\gc\phi/2}.\label{eq:kappa}
\end{align}

\subsection{Deviation equation for a ray}

Let us write
\begin{align}
    x^\mu=x_0^\mu(\lambda)+\expara \sigma\xi(\lambda),
\end{align}
where $x_0^\mu(\lambda)$ is a null geodesic that solves $D_0l_0^\mu = 0$.
Our goal is to determine $\xi^\mu$.
From Eq.~\eqref{SOeqs-l} we obtain the deviation equation~\cite{Frolov:2025bva}
\begin{align}
        D_0^2\xi^\mu +R_{\rho\nu\sigma}^{~~~~\mu}l_0^\rho l_0^\sigma\xi^\nu
        = \kappa^\mu.
\end{align}
Since $l^\mu l_\mu = 2\expara l_0^\mu D_0\xi_\mu = 0$, one can express
\begin{align}
        \xi^\mu&=\mathcal{X}_ll_0^\mu+\mathcal{X}_1e_1^\mu+\mathcal{X}_2e_2^\mu
        \notag \\ & =
        \mathcal{X}_ll_0^\mu+\bar{\mathcal{X}}\mathsf{m}_0^\mu+\mathcal{X}\bar{\mathsf{m}}_0^\mu,
        \label{eq:xi--calX}
\end{align}
where $\mathcal{X}_l(\lambda)$, $\mathcal{X}_1(\lambda)$, and $\mathcal{X}_2(\lambda)$ are real, $\mathsf{m}_0^\mu=(e_1^\mu+ie_2^\mu)/\sqrt{2}$, $\bar{\mathsf{m}}_0^\mu=(e_1^\mu-ie_2^\mu)/\sqrt{2}$, and $\mathcal{X}=(\mathcal{X}_1+i\mathcal{X}_2)/\sqrt{2}$.
Note that using a parallel-transported basis is more convenient here.
The right-hand side can be written in terms of the parallel-transported basis as
\begin{align}
    \kappa^\mu&=\left(\kappa_l+e^{-i\gc\phi/2}\ups\bar\kappa+e^{i\gc\phi/2}\bar\ups\kappa\right)l_0^\mu
    \notag \\ &\quad 
    +e^{-i\gc\phi/2}\bar\kappa \mathsf{m}_0^\mu 
    +e^{i\gc\phi/2}\kappa \bar{\mathsf{m}}_0^\mu,
\end{align}
yielding
\begin{align}
    &\frac{\D^2 \mathcal{X}_l}{\D \lambda^2}-
    R_{\rho\nu\sigma\mu}l_0^\rho\mathsf{m}_0^\nu l_0^\sigma\mathsf{n}_0^\mu \bar{\mathcal{X}}
    -
    R_{\rho\nu\sigma\mu}l_0^\rho\bar{\mathsf{m}}_0^\nu l_0^\sigma\mathsf{n}_0^\mu \mathcal{X}
    \notag \\
    &=\kappa_l+e^{-i\gc\phi/2}\ups\bar\kappa+e^{i\gc\phi/2}\bar\ups\kappa,
    \label{eq:ddXL}
    \\ 
    &\frac{\D^2 \mathcal{X}}{\D \lambda^2}+
    R_{\rho\nu\sigma\mu}l_0^\rho\mathsf{m}_0^\nu l_0^\sigma\mathsf{m}_0^\mu \bar{\mathcal{X}}
    +
    R_{\rho\nu\sigma\mu}l_0^\rho\bar{\mathsf{m}}_0^\nu l_0^\sigma\mathsf{m}_0^\mu \mathcal{X}
    \notag \\
    &=e^{i\gc\phi/2}\kappa.
    \label{eq:ddX}
\end{align}
In the rest of the paper, we solve these equations for a spherically symmetric configuration of the axion field in a spherically symmetric spacetime.
We also show in Appendix~\ref{app:FLRW} that there is no helicity-dependent bending of light due to spinoptics effects in a homogeneous and isotropic universe filled with ALPs.

\section{Scattering of light}\label{sec:examples}

As a demonstration, we consider the scattering of light by a Schwarzschild black hole dressed with an axion field.
The metric is given by
\begin{align}
    \D s^2= -f(r)\D{t^2}+ \frac{\D r^2}{f(r)}+ r^2\D \Omega^2,
\end{align}
where $f(r)= ( 1-2M/r)$ and $\D\Omega^2$ is a line element of a unit 2-sphere.

To obtain a complex null tetrad $(l_0^\mu,n_0^\mu,m_0^\mu,\bar m_0^\mu)$ associated to a null geodesic with the tangent vector $l_0^\mu$, we start with constructing a parallel-transported basis $(l_0,\mathsf{n}_0,\mathsf{m}_0,\bar{\mathsf{m}}_0)$ in the Schwarzschild spacetime.
This can be done following Ref.~\cite{Frolov:2024olb}.
(See Appendix~\ref{app:tetrad-for-spherical-sym} for a generalization.)
Without loss of generality, we may consider a null ray lying in the equatorial plane,
$\theta=\pi/2$.
Then, $l_0^\mu$ can be written as
\begin{align}
    l^\mu_0&=\left(
        \frac{E}{f},\frac{\mathcal{R}}{r},0,\frac{L}{r^2}
    \right),
    \\
    \mathcal{R}&=\pm\sqrt{E^2r^2-L^2f},
    \label{def:calR}
\end{align}
where $E$ and $L$ are the integrals of motion, with their physical meanings being obvious.
In the following numerical calculations, we set $E=1$.
The plus and minus signs in Eq.~\eqref{def:calR} correspond to outgoing and ingoing rays, respectively.
The turning point $r_{\textrm{m}}$ is defined by $\mathcal{R}(r_{\textrm{m}})=0$.
Using $r_{\textrm{m}}$, one can write $L/E=r_{\textrm{m}}/\sqrt{f(r_{\textrm{m}})}$.

We introduce
\begin{align}
    \tilde{e}_1^\mu&=\left(
        \frac{\mathcal{R}}{Lf},\frac{Er}{L},0,0
    \right),
    \\ 
    e_1^\mu&=\tilde{e}_1^\mu-\Phi l_0^\mu,
    \\ 
    e_2^\mu&=\left(
        0,0,-\frac{1}{r},0
    \right),
    \\
    \tilde{e}_3^\mu&=
    \left(
        \frac{Er^2}{2fL^2},\frac{r\mathcal{R}}{2L^2},0,-\frac{1}{2L}
    \right),
\end{align}
where $\Phi$ satisfies
\begin{align}
    D_0\Phi=\frac{E}{L}.
\end{align}
The general solution to this equation is given by
\begin{align}
    \Phi(\lambda)=\frac{E}{L}(\lambda+\lambda_0),
\end{align}
where $\lambda$ is an affine parameter and $\lambda_0$ is an integration constant.
Using these vectors, $\mathsf{m}_0^\mu$ and $\mathsf{n}_0^\mu$ can be constructed as
\begin{align}
    \mathsf{m}_0^\mu&=\frac{1}{\sqrt{2}}\left(e_1^\mu+ie_2^\mu\right),
    \\ 
    \mathsf{n}_0^\mu&=\tilde{e}_3^\mu-\Phi \tilde e_1^\mu+\frac{1}{2}\Phi^2 l_0^\mu.
\end{align}
We refer the readers to Ref.~\cite{Frolov:2024olb} for the detailed geometric description of the derivation of these results.
Having thus obtained the complex null tetrad $(l_0,\mathsf{n}_0,\mathsf{m}_0,\bar{\mathsf{m}}_0)$, we solve Eq.~\eqref{eq:Upsilon} to find $\ups$ and then construct $(l_0,n_0,m_0,\bar m_0)$ using Eqs.~\eqref{eq:def-m0} and~\eqref{eq:def-n0}.

The following results are useful for later calculations:
\begin{align}
    R_{\mu\nu\alpha\beta} l_0^\mu \mathsf{n}_0^\mu\bar{\mathsf{m}}_0^\alpha
    \mathsf{m}_0^\beta&=0,
    \\
    R_{\mu\nu\alpha\beta}\mathsf{m}_0^\mu l_0^\nu
                \bar{\mathsf{m}}_0^\alpha \mathsf{m}_0^\beta&=
                -\frac{3L^2M\Phi}{\sqrt{2}r^5},
    \\
    R_{\rho\nu\sigma\mu}l_0^\rho\mathsf{m}_0^\nu l_0^\sigma\mathsf{n}_0^\mu
    =
    R_{\rho\nu\sigma\mu}l_0^\rho\bar{\mathsf{m}}_0^\nu l_0^\sigma\mathsf{n}_0^\mu
    &=\frac{3L^2M\Phi}{\sqrt{2}r^5},
    \\ 
    R_{\rho\nu\sigma\mu}l_0^\rho\mathsf{m}_0^\nu l_0^\sigma\mathsf{m}_0^\mu
    &=-\frac{3L^2M}{r^5},
    \\ 
    R_{\rho\nu\sigma\mu}l_0^\rho\bar{\mathsf{m}}_0^\nu l_0^\sigma\mathsf{m}_0^\mu&=0.
\end{align}

We set $\lambda=0$ at the turning point $r=r_{\textrm{m}}$.
The affine parameter is then given by
\begin{align}
    \lambda=\int^r_{r_{\textrm{m}}}\frac{\rho}{\mathcal{R}(\rho)}\D \rho.
\end{align}
For $\lambda<0$, it is convenient to write this as
\begin{align}
    \lambda+c_{\textrm{m}}=-\frac{r}{E}+\int^\infty_r
    \left[\frac{\rho}{\sqrt{E^2\rho^2-L^2f(\rho)}}
    -\frac{1}{E}\right]\D \rho,
    \label{eq:lambda-rewrite}
\end{align}
where
\begin{align}
    c_{\textrm{m}}:=\int^{\infty}_{r_{\textrm{m}}}\left[\frac{\rho}{\sqrt{E^2\rho^2-L^2f(\rho)}}
    -\frac{1}{E}\right]\D \rho-\frac{r_{\textrm{m}}}{E}.
\end{align}
Note that the right-hand side of Eq.~\eqref{eq:lambda-rewrite} behaves as $-r/E+\mathcal{O}(r^{-1})$ for large $r$.
Now, by taking
\begin{align}
    \lambda_{0}=c_{\textrm{m}},
\end{align}
we find that $\Phi\approx -r/L+\mathcal{O}(r^{-1})$, and hence $e_1^\mu\approx (0,0,0,r^{-1})$, for incoming rays with negatively large $\lambda$.
For the right-handed polarization vector $M^\mu=m^\mu$ of incoming rays to coincide with $(e_1^\mu+ie_2^\mu)/\sqrt{2}$ up to an overall phase, we set $\lambda_0=c_{\textrm{m}}$ and $\ups_0=0$, which is a natural choice for a scattering problem~\cite{Frolov:2025bva}.

Using the expression~\eqref{eq:xi--calX}, one can write the spatial components of $\xi^\mu$ as
\begin{align}
    \xi^r&=\frac{\mathcal{R}}{r}\mathcal{X}_l+
    \left(\frac{Er}{L}-\Phi\frac{\mathcal{R}}{r}\right)\mathcal{X}_1,
    \label{eq:xir}
    \\ 
    \xi^\theta&=-\frac{\mathcal{X}_2}{r},
    \\ 
    \xi^\varphi&=\frac{L}{r^2}\left(\mathcal{X}_l-\Phi\mathcal{X}_1\right).
\end{align}
Assuming that the axion-dependent terms fall off sufficiently rapidly with increasing $r$, Eqs.~\eqref{eq:ddXL} and~\eqref{eq:ddX} become $\D^2\mathcal{X}_l/\D\lambda^2\approx 0$ and $\D^2\mathcal{X}/\D\lambda^2\approx 0$ for large $r$.
For positively large $\lambda$, we therefore have $\mathcal{X}_l/r\approx$ const and $\mathcal{X}/r\approx$ const.
Let us write the late-time asymptotics as
\begin{align}
    \frac{\mathcal{X}}{r}\approx -\frac{1}{\sqrt{2}}
    \left(\delta\varphi_\infty+i\delta\theta_\infty\right),
    \label{eq:calX--theta--phi}
\end{align}
where $\delta\varphi_\infty$ and $\delta\theta_\infty$ are real numbers.
Then, we get the asymptotic values as
\begin{align}
    \xi^\theta \approx \delta\theta_\infty ,
    \qquad 
    \xi^\varphi \approx \delta\varphi_\infty.
\end{align}

\subsection{Schwarzschild black hole without axion}\label{subsec:Sch-review}

Before discussing the impacts of the axion field, let us briefly review spinoptics in the Schwarzschild spacetime~\cite{Frolov:2024olb}, setting $\phi=0$.
In this case, Eqs.~\eqref{eq:ddXL} and~\eqref{eq:ddX} reduce to
\begin{align}
    &\frac{\D^2\mathcal{X}_l}{\D\lambda^2}
    -\frac{3L^2M\Phi}{r^5}\mathcal{X}_1=0,
    \\ 
    &\frac{\D^2\mathcal{X}}{\D\lambda^2}-\frac{3L^2M}{r^5}\bar{\mathcal{X}}
    =-i\frac{3L^2M\Phi}{\sqrt{2}r^5}.
\end{align}
We impose the initial conditions $\mathcal{X}_l=\mathcal{X}=\D\mathcal{X}_l/\D\lambda=\D\mathcal{X}/\D\lambda=0$ at $\lambda=-\infty$.
Then, the structure of the equations implies that $\mathcal{X}_l=\mathcal{X}_1=0$, and the only nonvanishing component is the imaginary part $\mathcal{X}_2$ of $\mathcal{X}$.
From Eq.~\eqref{eq:calX--theta--phi} we see that $\delta\varphi_\infty=0$ and $\delta\theta_\infty\neq 0$, i.e.,
the spinoptics effect in the Schwarzschild spacetime tilts the plane of the orbit while there is no shift in the $\varphi$ direction~\cite{Frolov:2024olb}.

\subsection{Axion in flat spacetime}

Let us first consider a spherically symmetric configuration of the axion field in a Minkowski spacetime by setting $M=0$.
This setup allows us to see how purely axionic effects come in.
We consider a light ray passing through a spherically symmetric axion dark-matter halo with the NFW profile~\cite{Navarro:1997AJ} with
\begin{align}
    \phi(t,r)&=\frac{\sqrt{2\rho_{\textrm{NFW}}(r)}}{m_\phi}\cos\left(m_\phi t\right),
    \label{eq:axion-nfw}
    \\ 
    \rho_{\textrm{NFW}}(r)&=\frac{\rho_s}{(r/r_s)(1+r/r_s)^2},
    \label{eq:rho-nfw}
\end{align}
where $m_\phi$, $r_s$, and $\rho_s$ are constant parameters.
Here, $m_\phi$ is the mass of ALPs.
In the present analysis, these parameters are chosen by hand to illustrate a qualitative behavior.

\begin{figure}
    \centering
    \includegraphics[width=1\linewidth]{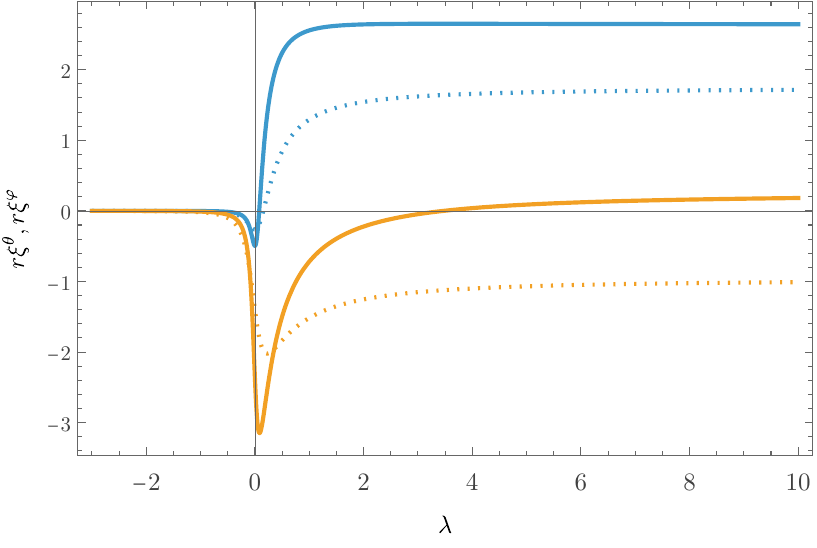}
    \caption{Displacements in the $y$ and $z$ directions ($r\xi^\theta$ and $r\xi^\varphi$) as functions of the affine parameter $\lambda$.
    The blue and orange lines correspond respectively to $ r\xi^\theta $and $r\xi^\varphi$, with $L=1/10$ (solid) and $L=1/5$ (dotted).}
    \label{fig:NFW_deviation}
\end{figure}

\begin{figure}
    \centering
    \includegraphics[width=1\linewidth]{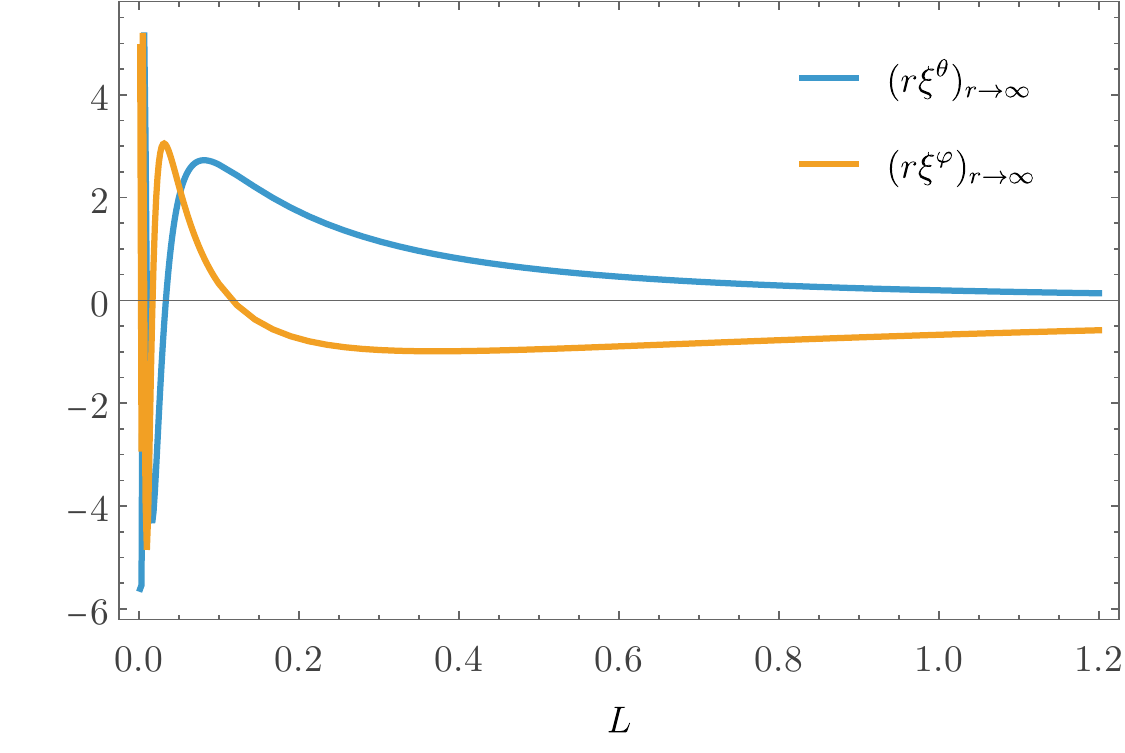}
    \caption{Asymptotic values of $r\xi^\theta$ and $r\xi^\varphi$ as functions of $L$.}
    \label{fig:nfw_L}
\end{figure}

\begin{figure}
    \centering
    \includegraphics[width=1 \linewidth]{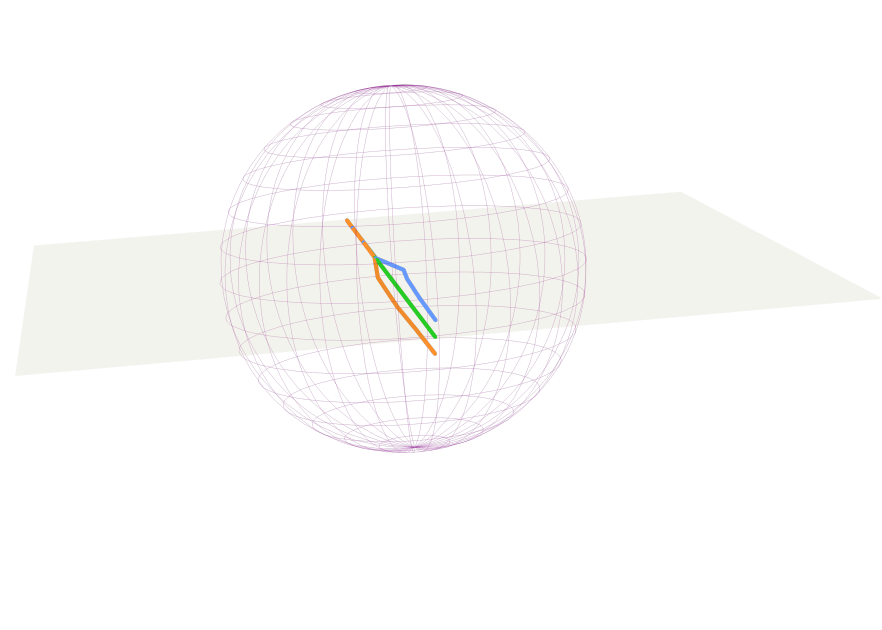}
    \caption{Ray trajectories passing through a spherical axion clump described by Eqs.~\eqref{eq:axion-nfw} and~\eqref{eq:rho-nfw} in Minkowski spacetime.
    The solid green curve represents a geodesic (i.e., a straight line), while the orange and blue curves represent trajectories obtained by taking into account the spinoptics effects from the axion--photon coupling, with the helicity parameter $\sigma=+1$ and $\sigma=-1$, respectively.
    For visualization purposes, we take $\expara=3\times 10^{-2}$.
    A sphere with the radius $r_s$ is also shown.}
    \label{fig:NFW_vis}
\end{figure}

We solve numerically Eqs.~\eqref{eq:ddXL} and~\eqref{eq:ddX} with $R_{\mu\nu\rho\sigma}=0$ and $\phi$ given above.
Rays are incident from $x=+\infty$ in the plane $z=0$ with impact parameter $L/E$.
The zeroth-order solution is a straight line with $y=L/E$ and $z=0$.
The late-time asymptotic behavior is written as
\begin{align}
    y&\approx \frac{L}{E}+\delta y,\qquad \delta y=-\expara \sigma (r\xi^\varphi)_{r\to\infty}
    =\textrm{const},
    \\ 
    z&\approx \delta z,\qquad \delta z= -\expara \sigma (r \xi^\theta)_{r\to\infty}
    =\textrm{const}.
\end{align}
Figure~\ref{fig:NFW_deviation} shows the solutions for $L=1/10$ and $L=1/5$.
The parameters of the axion profile are given by $m_\phi=1/10$, $r_s=1$, and $\rho_s=1/100$.
We see that rays are shifted in both the $y$ and $z$ directions.
The asymptotic values of $r\xi^\theta$ and $r\xi^\varphi$ are shown as functions of $L$ in Fig.~\ref{fig:nfw_L}.
Figure~\ref{fig:NFW_vis} visualizes ray trajectories with different helicity parameters for $L=1/10$.

\subsection{Axion around a Schwarzschild black hole}


Let us switch on gravity and consider a Schwarzschild black hole dressed with an axion field.
Ignoring the mass of $\phi$, the Klein--Gordon equation, $\Box\phi=0$, in the Schwarzschild spacetime admits the following solution:
\begin{align}
        \phi=\dot\phi_0 t+A_0\ln f(r),
\end{align}
where $\dot\phi_0$ and $A_0$ are constants.
The solution is regular at the future event horizon if $A_0=2M\dot\phi_0$,\footnote{This can be understood by using the ingoing Eddington--Finkelstein coordinate, $v=t+r+2M(r/2M-1)$.} 
\begin{align}
        \phi=\dot\phi_0\left[
                t+2M\ln f(r)
        \right].\label{eq:spherical_profile}
\end{align}

We now discuss the behavior of null rays in the axion profile \eqref{eq:spherical_profile}.
As reviewed in Sec.~\ref{subsec:Sch-review}, in the absence of the axion field, spinoptics corrections in Schwarzschild spacetime have been shown to appear only in the $\theta$ component, thereby tilting the orbital plane~\cite{Frolov:2024olb}. In the present case, however, the azimuthal deviation is expected to be non-vanishing, as implied by the analysis in the previous subsection.

\begin{figure}
    \centering
    \includegraphics[width=1\linewidth]{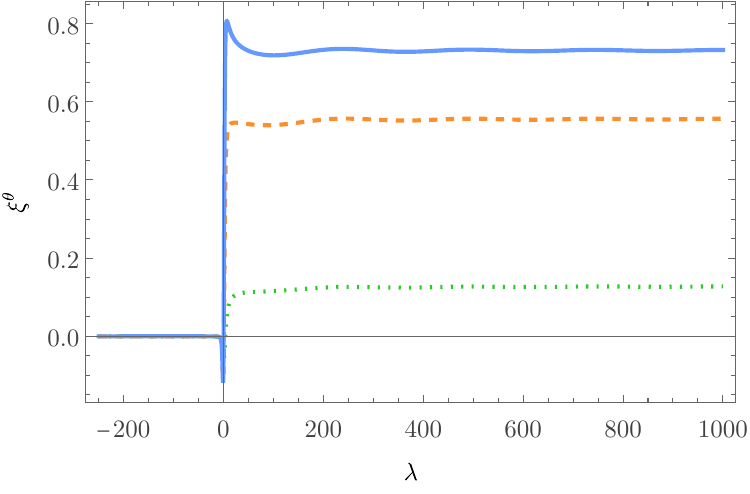}
    \includegraphics[width=1\linewidth]{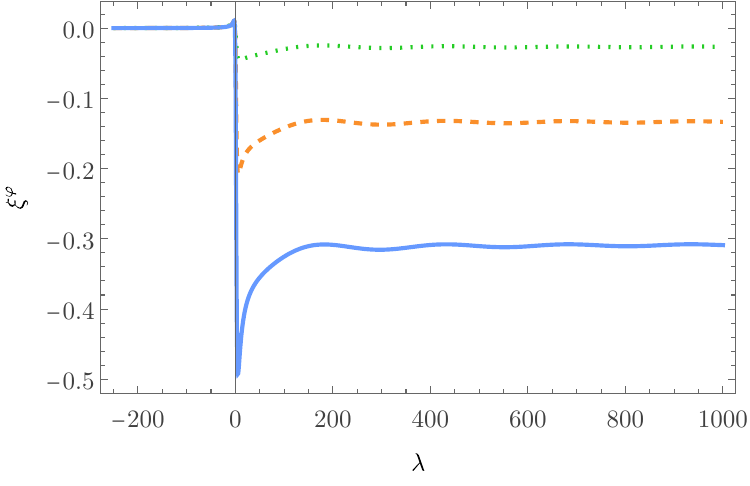}
    \caption{$\xi^\theta$ (upper panel) and $\xi^\varphi$ (lower panel) as functions of the affine parameter for $\gc\dot\phi_0=5\times 10^{-2}$.
    Solid blue, dashed orange, and dotted green lines correspond respectively to $r_\textrm{m}=1.8$, $r_\textrm{m}= 2.0$, and $r_\textrm{m}=3.0$.}
    \label{fig:axion--Sch_deviation}
\end{figure}

\begin{figure}
    \centering
    \includegraphics[width=1\linewidth]{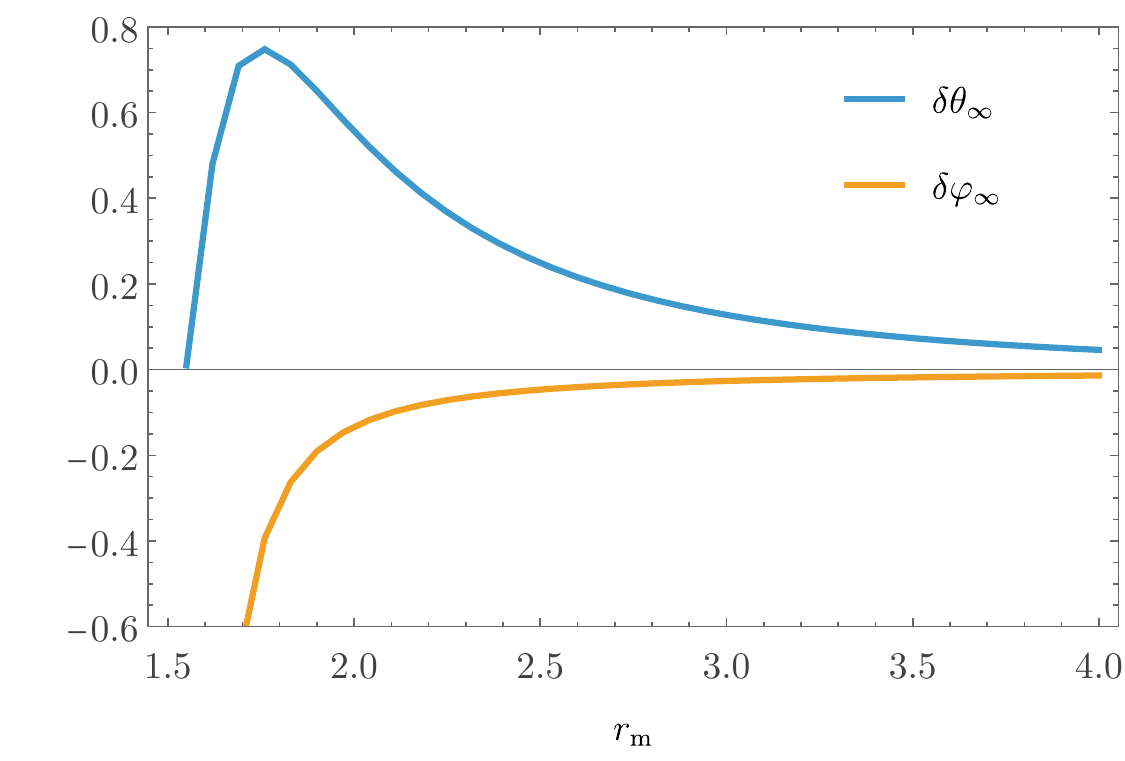}
    \caption{The asymptotic values $\delta\theta_\infty$ and $\delta\varphi_\infty$ as functions of $r_{\textrm{m}}$ for $\gc\dot\phi_0=5\times 10^{-2}$.} 
    \label{fig:bh_rm}
\end{figure}

\begin{figure}
    \centering
    \includegraphics[width=1\linewidth]{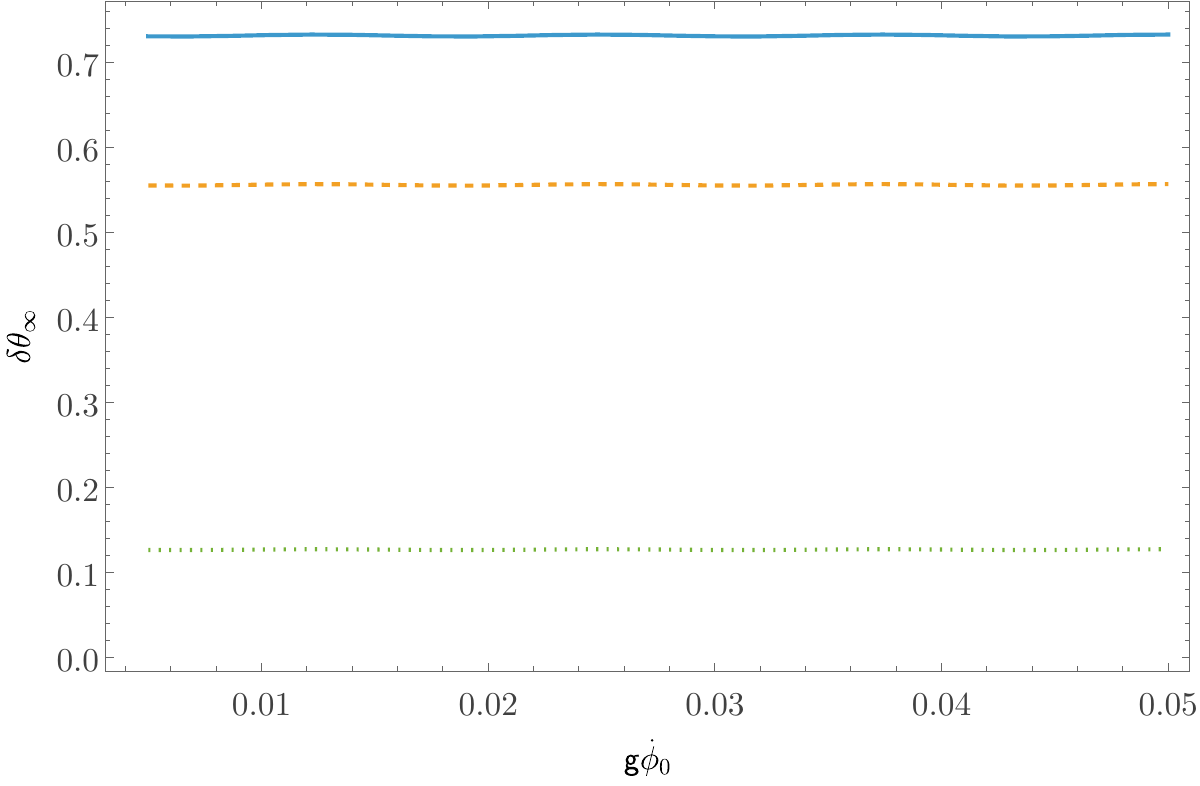}
    \includegraphics[width=1\linewidth]{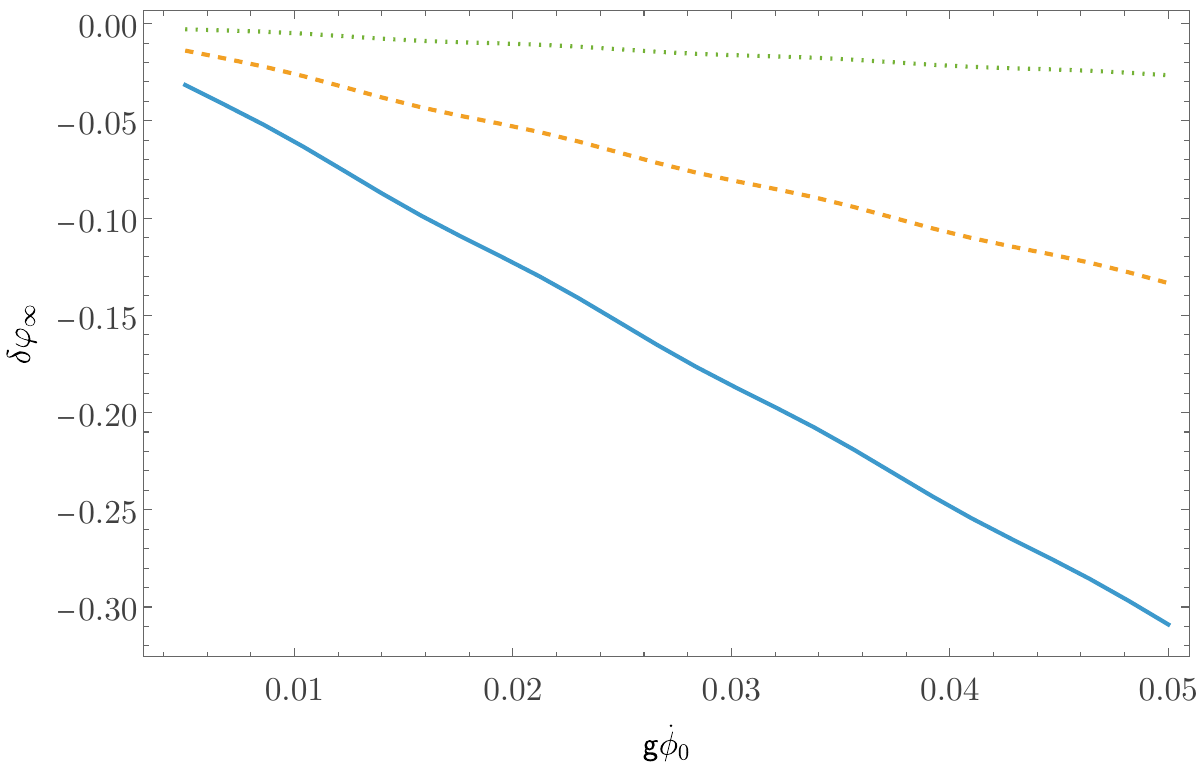}
    \caption{Asymptotic values $\delta\theta_\infty$ (upper panel) and $\delta\varphi_\infty$ (lower panel) as functions of $\gc\dot\phi_0$ for $r_\textrm{m}=1.8$ (sollid blue), $r_\textrm{m}=2.0$ (dashed orange), and $r_\textrm{m}=3.0$ (dotted green).
    }
    \label{fig:Sch_ALPs_asymp}
\end{figure}

\begin{figure}
    \centering
    \includegraphics[width=1\linewidth]{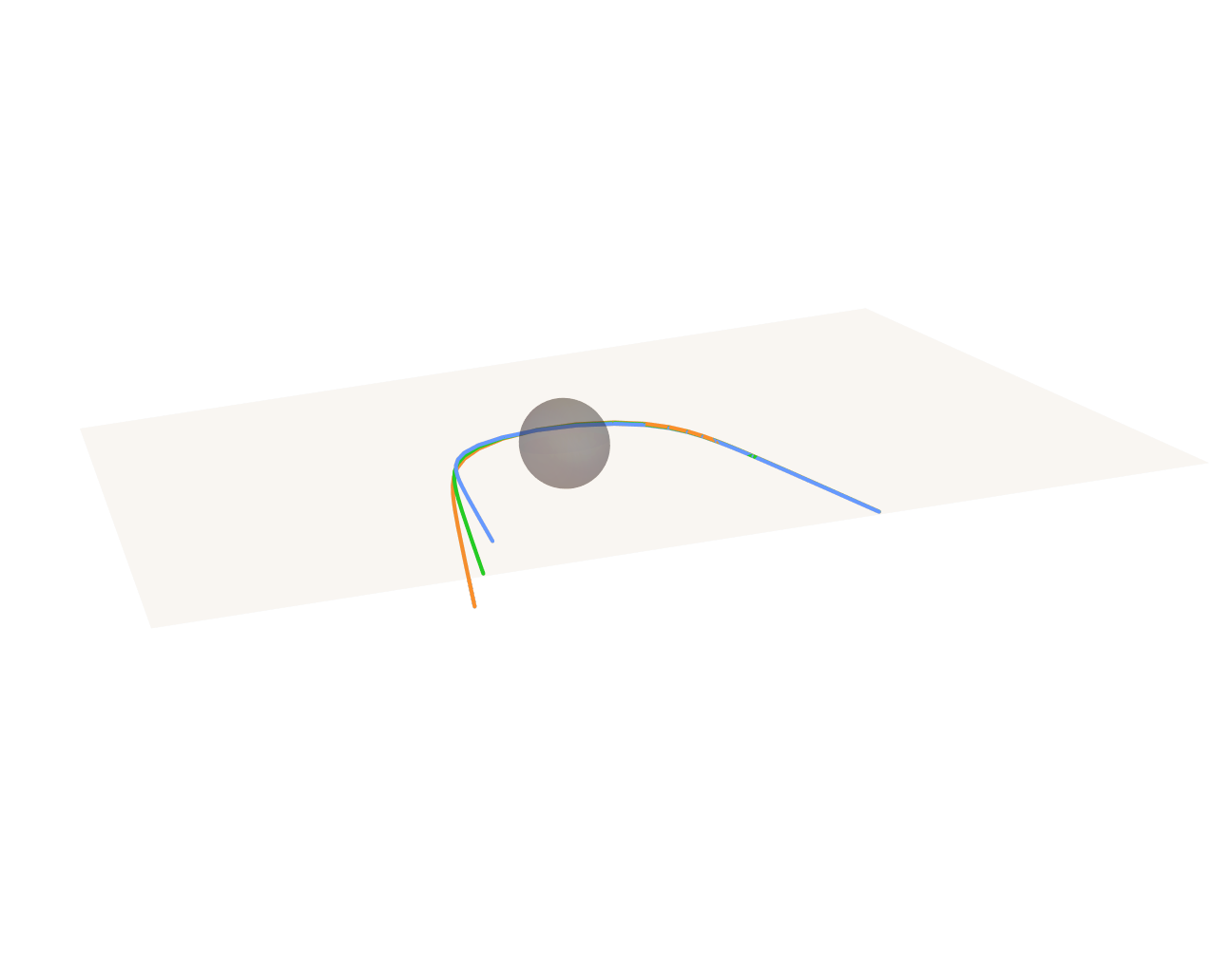}
    \caption{Ray trajectories passing by a Schwarzschild black hole with an axion hair~\eqref{eq:spherical_profile}.
    The green curve represents the geodesic, while the orange and blue curves represent the spinoptics-corrected trajectories incorporating the axion--photon coupling, for the helicity parameter $\sigma=+1$ and $\sigma=-1$, respectively.
    We set $r_\textrm{m}= 1.8$, and the other parameters are the same as those used in Fig.~\ref{fig:axion--Sch_deviation}. 
    For visualization purposes, we take $\expara=7\times10^{-2}$.
    The equatorial plane $\theta = \pi/2$ is shaded.
    The gray sphere represents the region $r\le 2M=1$}.
    \label{fig:axion-Sch}
\end{figure}

We set $M=1/2$.
Figure~\ref{fig:axion--Sch_deviation} shows $\xi^\theta$ and $\xi^\varphi$ for $\gc\dot \phi_0=5\times10^{-2}$ and different $r_{\textrm{m}}$.
It can be seen that $\xi^\theta$ and $\xi^\varphi$ exhibit damped oscillations and approach constant values.
These oscillations are due to the linear time-dependence of the axion field.
The asymptotic values depend on $r_{\textrm{m}}$ in a way that is shown in Fig.~\ref{fig:bh_rm}.
It is interesting to note that $\delta\theta_\infty$ is not a monotonic function of $r_{\textrm{m}}$.
In Fig.~\ref{fig:Sch_ALPs_asymp} we present the asymptotic values for different $r_{\textrm{m}}$ as functions of $\gc\dot\phi_0$.
We see that the asymptotic tilting angle $\delta\theta_\infty$ is insensitive to $\gc\dot\phi_0$, implying that this angle is determined by the gravitational effects rather than the axionic ones.
Figure~\ref{fig:axion-Sch} is a visualization of rays with different helicity parameters.

\section{Photon sphere}\label{sec:photon_sphere}

So far, we have studied the scattering of light by a black hole.
Let us turn to discuss how the notion of a photon sphere is affected by $\mathcal{O}(\expara)$ corrections in the spinoptics approximation.

Our zeroth-order orbit is now circular, with $r=3M$ and $L/E=3\sqrt{3}M$.
Since $\mathcal{R}=0$, it follows from Eq.~\eqref{eq:xir} that the radial deviation is determined solely from the real part of $\mathcal{X}$:
\begin{align}
    \xi^r=\frac{\mathcal{X}_1}{\sqrt{3}}.
\end{align}
Equations~\eqref{eq:ddXL} and~\eqref{eq:ddX} respectively read
\begin{align}
    \frac{\D^2\mathcal{X}_l}{\D\lambda^2}-\frac{E^2\Phi}{3M^2}\mathcal{X}_1&=
    \kappa_l+e^{-i\gc\phi/2}\ups\bar\kappa+e^{i\gc\phi/2}\bar\ups\kappa,
    \label{eq:circular-eq-XL}
    \\
    \frac{\D^2\mathcal{X}}{\D\lambda^2}-\frac{E^2}{3M^2}\bar{\mathcal{X}}&=e^{i\gc \phi/2}\kappa.
    \label{eq:circular-eq-X}
\end{align}

In the absence of the axion field, we have $\ups=0$, $\kappa_l=0$, and $\kappa=-(i/3\sqrt{2})(E^2/M^2)\Phi$, and the right-hand side of Eq.~\eqref{eq:circular-eq-X} is pure imaginary, as we have already seen in Sec.~\ref{subsec:Sch-review}.
In the present case we impose the initial conditions $\mathcal{X}_l=\mathcal{X}=\D\mathcal{X}_l/\D\lambda=\D\mathcal{X}/\D\lambda=0$ at $\lambda=0$.
The solution to Eqs.~\eqref{eq:circular-eq-XL} and~\eqref{eq:circular-eq-X} is then given by
\begin{align}
    \mathcal{X}_l&=\mathcal{X}_1=0,
    \\ 
    \mathcal{X}_2&=
    \frac{1}{3E}(\sin\varphi-\varphi),
\end{align}
where $\varphi=(E/\sqrt{3}M)\lambda$ and we set $\Phi(0)=0$.
(We have assumed that the zeroth-order orbit passes $\varphi=0$ at $\lambda=0$.)
This in particular shows that $\xi^r=0$, implying that light rays are trapped on the same photon sphere, even though they are no longer circular,
\begin{align}
    \xi^\theta=-\frac{\mathcal{X}_2}{3M}\neq 0,
\end{align}
once the $\mathcal{O}(\expara)$ corrections are taken into account.

In the presence of the axion field, it can be shown that the real part of the right-hand side of Eq.~\eqref{eq:circular-eq-X} is non-vanishing in general, yielding $\xi^r\neq 0$.
Suppose that $\phi$ depends only on $r$.
Equation~\eqref{eq:Upsilon} reads
\begin{align}
    \frac{\D\Xi}{\D\lambda}=i\frac{\gc\phi'(3M)}{2\sqrt{6}},
\end{align}
leading to the purely imaginary solution
\begin{align}
    \Xi(\lambda)= i\frac{\gc\phi'(3M)}{2\sqrt{6}}\lambda,
\end{align}
where we set $\Xi(0)=0$.
The right-hand side of Eq.~\eqref{eq:circular-eq-X} is now given by
\begin{align}
    e^{i\gc\phi}\kappa=\frac{E^2\textrm{Im}(\Xi)}{3M^2}-i\frac{E^2\Phi}{3\sqrt{2}M^2}.
\end{align}
The real part of the solution is given by
\begin{align}
    \mathcal{X}_1&=
    \frac{\gc\phi'(3M)M}{2E}\left(\sinh\varphi-\varphi\right),
\end{align}
which is indeed nonvanishing if $\phi'(3M)\neq 0$.
Thus, rays deviate away from the photon sphere due to the spinoptics effects.

It is straightforward to extend the discussion above to the case of a general static and spherically symmetric metric with $-g_{tt}\neq 1/g_{rr}$ by using a complex null tetrad derived in Appendix~\ref{app:tetrad-for-spherical-sym}.
In particular, in the absence of the axion field, Eq.~\eqref{eq:circular-eq-X} is modified to
\begin{align}
    \frac{\D^2\mathcal{X}}{\D\lambda^2}+\frac{L^2}{2r_0^4}
    \left(\mathcal{X}-\bar{\mathcal{X}}\right)
    -\frac{\mu_0^2}{2}\left(\mathcal{X}+\bar{\mathcal{X}}\right)
    =-i \frac{L^2\Phi}{\sqrt{2}r_0^4} ,
\end{align}
where $r_0$ is the radius of the zeroth-order circular orbit satisfying $r_0^2h'(r_0)/2h^2(r_0)=r_0^2/h(r_0)=L^2/E^2$ and $\mu_0^2:=L^2f(r_0)[1-L^2h''(r_0)/2E^2]/r_0^4$ with $g_{tt}=-h(r)$ and $g_{rr}=1/f(r)$.
Therefore, also in this case one has $\mathcal{X}_2\neq 0$ but $\xi^r\propto \mathcal{X}_1=0$: in a general static and spherically symmetric spacetime, light rays are trapped on the photon sphere in the absence of the axion field even when one takes into account the spinoptics effects.

\section{Conclusions}\label{sec:conclusions}

In this paper, we have investigated how the axion--photon coupling modifies ray trajectories in arbitrary curved spacetime beyond geometric optics limit.
For massless particles propagating in a gravitational field, previous studies have shown that finite-wavelength ``spinoptics'' corrections, obtained by extending the geometric optics approximation, affect their trajectories in a helicity-dependent way through the interaction between spacetime curvature and their internal spin degrees of freedom.
This phenomenon is known as the gravitational spin Hall effect.
We have extended the previous studies~\cite{Frolov:2011mh, Yoo:2012vv, Oancea:2019pgm, Frolov:2024ebe} to include the effects from the axion--photon coupling, providing a spinoptics prescription applicable to an arbitrary curved background and an arbitrary axion profile, as long as the characteristic wavelength of the Maxwell field is much smaller than the characteristic length scales over which the gravitational field and the axion field vary.
We have employed the effective action approach developed in Ref.~\cite{Frolov:2024ebe} to derive the spinoptics equations for the axion--Maxwell theory systematically.
In the geometric optics limit, it has been shown that a ray trajectory is a null geodesic even in the presence of an axion field~\cite{Blas:2019qqp, Fedderke:2019ajk, Schwarz:2020jjh}.
We have found that, in the spinoptics approximation, helicity-dependent deviations of a ray from a null geodesic arise not only from the curvature of spacetime but also from an axion field.

One of the key ingredients of our formulation is the construction of complex null tetrads associated with a null tangent vector to a ray.
Within the present approximation, it is sufficient to construct such tetrads at zeroth order, i.e., for a null geodesic.
We have introduced a parallel-transported tetrad and an ``axion-corrected'' tetrad, which are transformed into each other in a way that is related to axion-induced rotation of the polarization plane.
Both tetrads are crucial for our formulation.
We have demonstrated the construction of these tetrads for a static and spherically symmetric spacetime in the presence of an axion field that is dependent on the time and radial coordinates.


To illustrate examples, we have studied rays propagating through an axion clump in Minkowski spacetime and those scattered by a Schwarzschild black hole with an axion hair.
Following Ref.~\cite{Frolov:2025bva}, we have evaluated the angular components $\xi^\theta$ and $\xi^\varphi$ of a deviation vector from a null geodesic.
We have seen that deviations due to the spinoptics effects arise from the axion--photon interaction alone.
The spinoptics effects from gravitational fields give rise only to $\xi^\theta$, tilting the plane of the orbit.
In contrast, the axion-induced spinoptics effects yield both $\xi^\theta$ and $\xi^\varphi$, making a clear distinction between the gravitational and axionic effects.

We have also investigated the consequences of spinoptics corrections to a circular orbit of light.
We have found that, in the absence of an axion field, an orbit that is circular at zeroth order is no longer circular once the spinoptics effects are taken into account, but a ray still remains on a photon sphere.
In contrast, in the presence of (the radial gradient of) an axion field, the radial component of a deviation vector does not vanish, and hence a ray does not remain confined to a photon sphere.

While we have focused mainly on the spatial deviations of ray trajectories from null geodesics, it would be interesting to investigate the temporal deviations and the impacts of spinpotics on the arrival times of photons.
It would also be interesting to study how black hole shadows are modified due to the spinoptics effects.
In particular, since we have seen that the axion-induced spinoptics effects have significant impacts on the notion of a photon sphere, black hole shadows could be an invaluable probe for axion-like particles.
We leave these issues for future work.

\acknowledgments
We thank Tomohiro Harada, Takahisa Igata, Akihiro Ishibashi, Alex Koek, Hirotaka Yoshino, and especially Chul-Moon Yoo for many valuable discussions.
The work of TT was supported by the Rikkyo University Special Fund for Research.
The work of TK was supported by JSPS KAKENHI Grant No.~JP25K07308.

\appendix

\section{No helicity-dependent bending of light in an FLRW universe}\label{app:FLRW}

Let us show that in a Friedmann--Lema\^{i}tre--Robertson--Walker (FLRW) universe,
\begin{align}
    \D s^2=a^2(\eta)\left(-\D\eta^2+\D r^2+r^2\D\Omega^2\right),
\end{align}
and in the presence of a homogeneous axion field, $\phi=\phi(\eta)$, there is no helicity-dependent bending of light due to spinoptics effects (as expected).

A tangent to a null geodesic propagating along the radial direction is given by
\begin{align}
    l_0^\mu&=\left(\frac{1}{\sqrt{2}a^2},\frac{1}{\sqrt{2}a^2},0,0\right),
\end{align}
and a parallel-transported complex null tetrad associated to it is obtained as
\begin{align}
    \mathsf{n}_0^\mu&=\left(\frac{1}{\sqrt{2}},-\frac{1}{\sqrt{2}},0,0\right),
    \\
    \mathsf{m}_0^\mu&=\left(0,0,\frac{1}{\sqrt{2}ar},-\frac{i}{\sqrt{2}ar}\right).
\end{align}
Equation~\eqref{eq:Upsilon} reads $D_0\left(e^{-i\gc\phi/2}\ups\right)=0$, and hence we choose $\ups=0$.
An axion-corrected basis is therefore given by $n^\mu_0=\mathsf{n}_0^\mu$ and $m_0^\mu=e^{-i\gc \phi/2}\mathsf{m}_0^\mu$.
We have
\begin{align}
    \kappa_l=\frac{\gc}{2a^2}\frac{\D^2\phi}{\D\eta^2},
    \qquad 
    \kappa=0,
\end{align}
showing that $\mathcal{X}=0$ and $\mathcal{X}_l\neq 0$. 
The deviation vector has only a tangential component, $\xi^\mu=\mathcal{X}_ll_0^\mu$, and thus we conclude that no helicity-dependent bending of light occurs in an FLRW universe.

\section{Complex null tetrad in general static and spherically symmetric spacetime}\label{app:tetrad-for-spherical-sym}

In Sec.~\ref{sec:examples}, we only consider rays in the Schwarzschild spacetime.
In this Appendix, we discuss null rays and parallel-transported complex null tetrads in the general static and spherically symmetric spacetime with the metric
\begin{align}
    \D s^2=-h(r)\D t^2+\frac{\D r^2}{f(r)}+r^2\D\Omega^2,\label{metric:gen}
\end{align}
generalizing the results of Ref.~\cite{Frolov:2024olb} to the case with $f\neq h$.
In the final stage of this work, we became aware of a very recent work~\cite{Alves:2026jyc}, in which the same results have been derived.

The tangent vector $l^\mu$ of a null geodesic in the spacetime with the metric~\eqref{metric:gen} can be written as
\begin{align}
    l^\mu =\left(\frac{E}{h},\frac{\mathcal{R}}{r},0,\frac{L}{r^2}\right),
\end{align}
where $E$ and $L$ are integrals of motion, and
\begin{align}
    \mathcal{R}=\pm\sqrt{\frac{f}{h}\left(E^2r^2-L^2h\right)}.
\end{align}
Here, we assumed without loss of generality that the null geodesic lies in the equatorial plane, $\theta=\pi/2$.

A complex null tetrad $(l^\mu,\mathsf{n}^\mu, \mathsf{m}^\mu,\bar{\mathsf{m}}^\mu)$ parallel-propagated along the null geodesic can be obtained as follows.
First, we introduce $\tilde e_1^\mu$ and $e_2$ defined as
\begin{align}
    \tilde e_1^\mu&=\left(
        \frac{\mathcal{R}}{L\sqrt{fh}},\sqrt{\frac{f}{h}}\frac{Er}{L},0,0
    \right),
    \\ 
    e_2^\mu&=\left(0,0,-\frac{1}{r},0\right).
\end{align}
These vectors satisfy
\begin{align}
    &(\tilde e_1,\tilde e_1)=(e_2,e_2)=1,
    \\
    &(\tilde e_1,e_2)=(\tilde e_1,l)=(e_2,l)=0.
\end{align}
Although $e_2^\mu$ is parallel-propagated along the null geodesic ($l^\nu\nabla_\nu e_2^\mu=0$), $\tilde e_1$ is not.
We try a construction analogous to Ref.~\cite{Frolov:2024olb} and consider the following ``improved'' vector:
\begin{align}
    e_1^\mu=\tilde e_1^\mu-\Phi(\lambda)l^\mu.
\end{align}
One sees that $e_1^\mu$ is parallel-propagated along the null geodesic if $\Phi$ satisfies
\begin{align}
    \frac{\D\Phi}{\D\lambda}=\frac{E}{L}\sqrt{\frac{f}{h}}.
    \label{eq:Phi:gen}
\end{align}
One can then define $\mathsf{m}^\mu$ as
\begin{align}
    \mathsf{m}^\mu=\frac{1}{\sqrt{2}}\left(e_1^\mu+ie_2^\mu\right).
\end{align}
We also introduce the vector $\tilde e_3^\mu$ defined as
\begin{align}
    \tilde e_3^\mu=\left(\frac{Er^2}{2L^2h},\frac{r\mathcal{R}}{2L^2},0,-\frac{1}{2L}\right),
\end{align}
which satisfies
\begin{align}
    (\tilde e_3,l)=-1,\qquad (\tilde e_1, \tilde e_3)=(e_2,\tilde e_3)=0.
\end{align}
Using $\tilde e_3^\mu$, one can define 
\begin{align}
    \mathsf{n}^\mu=\tilde e_3^\mu-\Phi\tilde e_1^\mu+\frac{1}{2}\Phi^2l^\mu,
\end{align}
which is parallel-propagated along the null geodesic.
It is easy to check that the vectors $(l^\mu,\mathsf{n}^\mu, \mathsf{m}^\mu,\bar{\mathsf{m}}^\mu)$ satisfy the required normalization conditions.

In the simplest case with $f=h$, Eq.~\eqref{eq:Phi:gen} can be integrated immediately to give $\Phi=(E/L)(\lambda+\lambda_0)$.
In general, we need to integrate Eq.~\eqref{eq:Phi:gen} numerically.
Let us now discuss the appropriate initial condition for $\Phi$ in the case of a scattering problem.
One can write
\begin{align}
    \Phi&=\frac{E}{L}\int_0^\lambda\sqrt{\frac{f}{h}}\D\lambda'+\Phi_{\textrm{m}}
    \notag \\
    &=\frac{E}{L}\int^r_{r_{\textrm{m}}}\sqrt{\frac{f}{h}}\frac{\rho}{\mathcal{R}}\D\rho+\Phi_{\textrm{m}},
\end{align}
where $\Phi_{\textrm{m}}$ is an integration constant and we have set $\lambda=0$
at the turning point $r=r_{\textrm{m}}$.
For a light ray coming in from infinity toward the turning point, $\Phi$ can be written as
\begin{align}
    \Phi&=-\frac{E}{L}\int^r_{r_{\textrm{m}}}\frac{\rho\D\rho}{\sqrt{E^2\rho^2-L^2h}}+\Phi_{\textrm{m}}
    \notag \\ &=
    -\frac{r}{L}+\frac{E}{L}\int^\infty_r\left(
                \frac{\rho}{\sqrt{E^2\rho^2-L^2h}}-\frac{1}{E}
        \right)\D\rho
        \notag \\ & \quad
        -\frac{E}{L}c_{\textrm{m}}+\Phi_{\textrm{m}},
\end{align}
where
\begin{align}
    c_{\textrm{m}}:=\int^\infty_{r_\textrm{m}}\left(
                \frac{\rho}{\sqrt{E^2\rho^2-L^2h}}-\frac{1}{E}
        \right)\D\rho-\frac{r_{\textrm{m}}}{E}.
\end{align}
For large $r$, $\Phi$ behaves as $\Phi=-r/L-(E/L)c_{\textrm{m}}+\Phi_{\textrm{m}}+\mathcal{O}(r^{-1})$,
and hence
$e_1^t=-E[\Phi_{\textrm{m}}-(E/L)c_{\textrm{m}}]+\mathcal{O}(r^{-1})$
and
$e_1^r=E[\Phi_{\textrm{m}}-(E/L)c_{\textrm{m}}]+\mathcal{O}(r^{-1})$.
Thus, by choosing the integration constant as
\begin{align}
    \Phi_{\textrm{m}}=\frac{E}{L}c_{\textrm{m}},
\end{align}
we have $e_1^\mu\approx (0,0,0,r^{-1})$ for large $r$, implying that
$e_1^\mu\approx e_\varphi^\mu$ and $e_2^\mu\approx -e_\theta^\mu$ initially.

\bibliography{refs}
\bibliographystyle{JHEP}

\end{document}